\documentclass{article}

\usepackage{wrapfig}
\usepackage{amsmath}
\usepackage{amssymb}
\usepackage{fullpage}
\usepackage{enumerate}
\usepackage{graphicx}
\usepackage{subcaption}
\usepackage{ifthen}
\usepackage{xcolor}
\usepackage{authblk}

\usepackage[nottoc]{tocbibind}

\usepackage[a4paper,top=3cm,bottom=2cm,left=3cm,right=3cm,marginparwidth=1.75cm]{geometry}
\title{Analysis of fine-tuning measures in models with extended Higgs sectors}

\author[1]{Dani\"el Boer\thanks{d.boer@rug.nl}}

\author[1]{Ruud Peeters\thanks{r.j.c.peeters@rug.nl}}

\author[2]{Sybrand Zeinstra\thanks{swzeinstra@uni-muenster.de}}

\affil[1]{Van Swinderen Institute for Particle Physics and Gravity, University of Groningen, Nijenborgh 4, 9747 AG Groningen, The Netherlands}
\affil[2]{Institut f\"ur Theoretische Physik, Westf\"alische Wilhelms-Universit\"at M\"unster, Wilhelm-Klemm-Stra{\ss}e 9, 48149, M\"unster, Germany}

\date{}

\newcommand{\order}[1]{$\mathcal{O}(#1)$}

\newcommand{\phif}[2]{\ifthenelse{\equal{#1}{1}}
	{
		\ifthenelse{\equal{#2}{1}}
		{\tilde{\phi}^{\dagger}}
		{\phi^{\dagger}}
	}
	{
		\ifthenelse{\equal{#2}{1}}
		{\tilde{\phi}}
		{\phi}
	}
}

\newcommand{\Deltaf}[2]{\ifthenelse{\equal{#2}{1}}
	{\Delta_{#1}^{\dagger}}
	{\Delta_{#1}^{\phantom{\dagger}}}
}

\newcommand{\tr}[1]{\text{Tr}(#1)}

\newcommand{\Aone}{\tr{\phif{0}{0}\phif{1}{0}}}
\newcommand{\Atwo}{\tr{\phif{0}{0}\phif{1}{1}}}
\newcommand{\Athree}{\tr{\phif{0}{1}\phif{1}{0}}}

\newcommand{\AL}{\tr{\Deltaf{L}{0}\Deltaf{L}{1}}}
\newcommand{\AR}{\tr{\Deltaf{R}{0}\Deltaf{R}{1}}}

\newcommand{\BLL}{\tr{\Deltaf{L}{0}\Deltaf{L}{0}}\tr{\Deltaf{L}{1}\Deltaf{L}{1}}}
\newcommand{\BRR}{\tr{\Deltaf{R}{0}\Deltaf{R}{0}}\tr{\Deltaf{R}{1}\Deltaf{R}{1}}}
\newcommand{\BLR}{\tr{\Deltaf{L}{0}\Deltaf{L}{0}}\tr{\Deltaf{R}{1}\Deltaf{R}{1}}}
\newcommand{\BRL}{\tr{\Deltaf{R}{0}\Deltaf{R}{0}}\tr{\Deltaf{L}{1}\Deltaf{L}{1}}}

\newcommand{\CL}{\tr{\phif{1}{0}\Deltaf{L}{0}\Deltaf{L}{1}\phif{0}{0}}}
\newcommand{\CR}{\tr{\phif{0}{0}\Deltaf{R}{0}\Deltaf{R}{1}\phif{1}{0}}}

\newcommand{\Done}{\tr{\phif{1}{0}\Deltaf{L}{0}\phif{0}{0}\Deltaf{R}{1}}}
\newcommand{\Dtwo}{\tr{\phif{1}{0}\Deltaf{L}{0}\phif{0}{1}\Deltaf{R}{1}}}
\newcommand{\Dthree}{\tr{\phif{1}{1}\Deltaf{L}{0}\phif{0}{0}\Deltaf{R}{1}}}
\newcommand{\Dones}{\tr{\phif{1}{0}\Deltaf{L}{1}\phif{0}{0}\Deltaf{R}{0}}}
\newcommand{\Dtwos}{\tr{\phif{1}{1}\Deltaf{L}{1}\phif{0}{0}\Deltaf{R}{0}}}
\newcommand{\Dthrees}{\tr{\phif{1}{0}\Deltaf{L}{1}\phif{0}{1}\Deltaf{R}{0}}}

\usepackage[normalem]{ulem} 
\renewcommand\sout{\bgroup \color[rgb]{0.55,0.00,0.99} \ULdepth=-.5ex \ULset}

\begin{document}
\maketitle

\begin{abstract}
In the literature measures of fine-tuning have been discussed as one of the tools to assess the feasibility of beyond the Standard Model theories. In this paper we focus on two specific measures and investigate what kind of fine-tuning they actually quantify. First we apply both measures to the two Higgs doublet model, for which one can analyze the numerical results in terms of available analytic expressions. After drawing various conclusions about the fine-tuning measures, we investigate a particular left-right symmetric model for which it has been claimed that already at tree-level it suffers from a high amount of fine-tuning. We will reach a different conclusion, although left-right symmetric models may require a modest amount of fine-tuning if phenomenological constraints are imposed. Our analysis shows that the two considered measures can probe different aspects of fine-tuning and are both useful if applied and interpreted in the appropriate way. 
\end{abstract}

\section{Introduction}
In the search for new physics beyond the Standard Model of elementary particles, there are no experimental indications to guide the theoretical modelling in a particular direction. There are however stringent bounds on the masses of new particles and on some new interactions that would violate symmetries of the Standard Model (SM), which set constraints on the particle content of the theory and on the possible interactions between them. Furthermore, in the construction of theories beyond the SM it is common to follow general guiding principles, such as the degree of symmetry and naturalness of the theory. Naturalness means that the theory should not possess extraordinarily small parameters, unless for symmetry reasons. Moreover, very large parameters have the problem that they prevent perturbative analyses which, apart from the strong interactions at low energies, have been very successful in the development of the SM. Imposing naturalness and perturbativity requirements on a theory is arguable, but commonly considered in theoretical investigations of beyond the SM physics in order to limit the vast amount of possibilities to some extent. Another requirement on the parameters of a theory that is often considered is that they should not be fine-tuned, i.e.\ no large cancellations between independent quantities should be needed. Although this might seem like an obvious requirement, it is in practice not straightforward to quantify fine-tuning, nor to decide what constitutes a too large amount of fine-tuning or how to ensure that no fine-tuning arises in a theory to begin with. In the literature this topic has mostly been investigated in the context of supersymmetric extensions of the SM (SSM), see e.g.\ \cite{Ross:2017kjc}, where the Higgs mass of 125 GeV puts tension on many restricted forms of SSMs. In these studies the measure put forward by Barbieri and Guidice \cite{Barbieri:1987fn} is employed, which quantifies the dependence of masses of particles on the parameters in the theory. Many simplified SSMs already display fine-tuning at the level of 3 significant digits, but there are still versions that are without fine-tuning, see e.g.\ \cite{vanBeekveld:2016hug,Baer:2013gva}. 

Another measure of fine-tuning inspired by the Barbieri-Guidice measure has been considered in the study of left-right symmetric models \cite{Dekens:2014ina}. In this measure, the relations between the parameters in the scalar potential are considered in all possible variations, using the equations that determine the minimum of the potential. A conclusion of  \cite{Dekens:2014ina} was that several popular left-right symmetric models display a very large amount of fine-tuning. In contrast, we will reach a different conclusion in this paper. In order to learn about the meaning of both these measures and about the significance of these measures being large, we have first performed an analysis of the two Higgs doublet model (2HDM), which has the advantage that all investigations can be done numerically as well as analytically. Moreover, the model is similar to the Higgs sector of the constrained minimal SSM (cMSSM) and has similar features as the Higgs sector in the left-right symmetric models (LRSMs), to which we turn subsequently. One conclusion will be that neither the 2HDM nor the LRSM have a large amount of fine-tuning, despite a possibly large hierarchy of scales, except if one imposes certain constraints on masses of scalars. Although our analysis will have some resemblance to the hierarchy problem it is restricted purely to the classical level. We therefore do not aim to shed new light on the hierarchy problem (for a study of fine-tuning beyond tree level in the 2HDM see for instance \cite{Casas:2006bd}). Another conclusion will be that both measures can provide information on fine-tuning but in some cases they concern complementary aspects of the theory. Therefore, neither measure fully specifies or captures the possible fine-tuning that may be present in a theory. 

The article is set up as follows: in Section 2 we discuss the two fine-tuning measures in more detail. These measures will then be applied to the two Higgs doublet model in Section 3. The results from this analysis will then be applied to the more complex left-right symmetric model in Section 4. We will conclude in Section 5.

\section{Fine-tuning measures} \label{sec:measures}
Before drawing conclusions about fine-tuning in beyond the Standard Model theories, one first needs a way to quantify the amount of fine-tuning in a theory, through a fine-tuning measure typically denoted by $\Delta$. Such a measure produces a number characterizing the amount of fine-tuning in a model point of a theory, i.e.\ for specific values of the model parameters. One can roughly interpret the logarithm of $\Delta$ as the amount of significant digits that need to be tuned in at least one of the parameters of the theory. So a value of $\Delta$ = 1000 corresponds to having at least one parameter that needs to be tuned up to three digits. We say roughly because it matters whether one considers the dependence on the parameters of a mass or of a mass squared for instance. The distinction is not relevant and therefore we discuss orders of magnitude only.

In this paper we will investigate fine-tuning in the Higgs sector in two different ways. First of all, we will look at the (tree-level) minimum equations of the scalar potential. We solve these equations for a set of parameters to obtain the values of these parameters. Then we will use the Dekens measure \cite{Dekens:2014ina} to identify the amount of fine-tuning in these equations. When using the Dekens measure, we split the set of parameters in two groups. We have a set of dependent parameters $q_i$, which depend on the remaining independent parameters of the theory $p_j$. The dependent parameters can in principle be different from the ones that were solved for through the minimum equations. After rewriting the minimum equations in terms of the parameters $q_i$ (while still solving for the same parameters), we determine the amount of fine-tuning by taking logarithmic derivatives of the $q_i$ with respect to the $p_j$.
\begin{equation}
\Delta_D = \max_{i,j}\ \lvert \Delta_D(q_i,p_j)\rvert = \max_{i,j} \left\lvert\frac{\partial \log q_i}{\partial \log p_j}\right\rvert = \max_{i,j}\left\lvert\frac{p_j}{q_i}\frac{\partial q_i}{\partial p_j}\right\rvert.
\end{equation}
The logarithmic derivative ensures that not only the sizes of parameters are compared, but also the magnitudes of the changes in the parameters. 
This means that if a change in a parameter $p_i$ by one unit results in a change of another parameter by a factor of $10^{3}$, the measure will yield a large value due to the large sensitivity of one parameter on the other. This would be considered a sign of fine-tuning. 

The second way we will investigate the amount of fine-tuning is by looking at observables, where in our analysis we consider particle masses. This will be done by using the Barbieri-Giudice (BG) measure \cite{Barbieri:1987fn}. This measure is similar in form to the Dekens measure, but it compares parameters to observables, and not to other parameters. If we have a set of observables $O_i$ which depend on a set of parameters $p_j$, the BG measure is defined as:
\begin{equation}
\Delta_{BG} = \max_{i,j} \left\lvert\frac{\partial \log O_i}{\partial \log p_j}\right\rvert = \max_{i,j} \left\lvert \frac{p_j}{O_i}\frac{\partial O_i}{\partial p_j}\right\rvert
\end{equation}

In the following sections we will take a close look at how these measures work and which kinds of fine-tuning they measure. This will lead to various insights on how to apply the measures, and on the amount of fine-tuning in theories with a large hierarchy.

\section{Fine-tuning study of the Two Higgs Doublet Model}
The Two Higgs Doublet Model (2HDM) is one of the simpler extensions of the Standard Model. Its relative simplicity makes it possible to find analytic expressions for the quantities of interest, while having a rich enough structure to be interesting from a fine-tuning perspective, provided by a large hierarchy of scales. Before analyzing the different fine-tuning measures in this model, we will discuss the essentials of the model.

\subsection{The Higgs potential of the 2HDM}
The Higgs sector of the 2HDM is constructed by adding an additional doublet to the Standard Model Higgs sector \cite{Branco:2011iw}. The two doublets are defined as:
\begin{equation}
  \Phi_1 = \begin{pmatrix} \phi_1^+ \\ \phi_1^0 \\
\end{pmatrix},
\quad
\Phi_2 = \begin{pmatrix} \phi_2^+ \\ \phi_2^0 \\
\end{pmatrix}.
\end{equation}
In order to simplify our discussion, we demand $CP$ invariance and impose a $\mathbb{Z}_2$ symmetry on the potential. Under these constraints, the Higgs potential has the form:
\begin{align*}
V = -\mu_1^2 A - \mu_2^2 B +\lambda_1A^2 + \lambda_2B^2 +\lambda_3C^2+\lambda_4D^2+\lambda_5AB,
\end{align*}
where the invariants $A, B, C$ and $D$ are defined as:
\begin{align*}
  A &= \Phi_1^{\dagger}\Phi_1, \\
  B &= \Phi_2^{\dagger}\Phi_2, \\
  C &= \frac{1}{2}\left(\Phi_1^{\dagger}\Phi_2 + \Phi_2^{\dagger}\Phi_1\right), \\
  D &= \frac{1}{2i}\left(\Phi_1^{\dagger}\Phi_2 - \Phi_2^{\dagger}\Phi_1\right).
\end{align*}
We are interested in the minimum of this potential, so we need to introduce
vacuum expectation values (vevs) for the two doublets. We will set:
\begin{equation}
  \langle\Phi_1\rangle = \frac{1}{\sqrt{2}}\begin{pmatrix} 0 \\ v_1 \\
\end{pmatrix}
, \langle\Phi_2\rangle = \frac{1}{\sqrt{2}}\begin{pmatrix} 0 \\ v_2 \\
\end{pmatrix}.
\end{equation}
In the standard 2HDM, the two vevs have to satisfy the relation $v_1^2 + v_2^2 = v^2 = (246 \text{ GeV})^2$. However, we will not impose this constraint, because we are not aiming to do phenomenology with this model. For the sake of studying fine-tuning we will consider the model with a large hierarchy of scales, i.e.\ we will impose that $v_1 \gg v_2$. This corresponds to taking the decoupling limit. The results are independent of the actual value chosen for $v$, which we will set to the arbitrarily chosen value of 50 TeV.

After taking derivatives with respect to the fields and inserting the vevs we obtain the minimum equations:
\begin{align}
\begin{split}
v_1(-\mu_1^2 + \lambda_1v_1^2 + \lambda_+ v_2^2) &= 0 \\
v_2(-\mu_2^2 + \lambda_2v_2^2 + \lambda_+ v_1^2) &= 0,
\end{split}
\end{align}
where $\lambda_+$ is defined as: $\lambda_+ = \frac{1}{2}(\lambda_3+\lambda_5)$. Assuming that both $v_1$ and $v_2$ are non-zero\footnote{In the Higgs basis where one rotates one of the vevs to zero, the same results are obtained.}, we can write:
\begin{align}
	\begin{split}
	\mu_1^2 = \lambda_1 v_1^2 + \lambda_+ v_2^2, \\
	\mu_2^2 = \lambda_2 v_2^2 + \lambda_+ v_1^2.
	\label{eq:minEquations}
	\end{split}
\end{align}
We can solve this set of equations for any set of two parameters, but we choose to solve for the two $\mu^2$ parameters. This makes sense in our case since we want to impose constraints on all the other parameters. We want the coupling constants to be of \order{1} (perturbative and natural, which in practice means we will consider values in the range $[0.1,10]$), and we want to impose a hierarchy on the vevs. By solving the minimum equations in this way and imposing the constraints mentioned above, we see that both $\mu^2$ parameters will in general be of order $v_1^2$. So both mass parameters are insensitive to the hierarchy in scales, they will both be of the order of the highest scale in the theory.

\subsubsection{Masses}
Due to the fact that there are two doublets instead of one, the particle content of the 2HDM is richer than in the Standard Model. There are now five scalar bosons after electroweak symmetry breaking. For our purposes, the masses of the $CP$-even states $h$ and $H$ are the most relevant: 
\begin{equation}
m_{h,H}^2 = \lambda_1v_1^2 + \lambda_2v_2^2 \mp \sqrt{(\lambda_1v_1^2-\lambda_2v_2^2)^2+4\lambda_+^2v_1^2v_2^2}.
\end{equation}
Since we are using that $v_1 \gg v_2$, we can write approximate expressions for these masses:
\begin{align}
m_h^2 &\approx 2\lambda_2v_2^2 -\frac{(\lambda_3+\lambda_5)^2}{2\lambda_1}v_2^2, \\
m_H^2 &\approx 2\lambda_1v_1^2 +\frac{(\lambda_3+\lambda_5)^2}{2\lambda_1}v_2^2.
\end{align}
So we see that the lightest Higgs will be naturally light, while the heavier Higgs will be of the order of the high scale.

\subsubsection{Boundedness of the potential}
There are two sets of constraints that need to be satisfied by the potential. First of all, the potential must be bounded from below. This can be ensured by demanding (see e.g.\ \cite{Branco:2011iw,Ferreira:2009jb,Ivanov:2006yq}):
\begin{align}
\begin{split}
\lambda_1 &> 0 ,\\
\lambda_2 &> 0, \\
\lambda_5 &> -2\sqrt{\lambda_1\lambda_2}, \\
\lambda_3 + \lambda_5 &> -2\sqrt{\lambda_1\lambda_2}, \\
\lambda_4 + \lambda_5 &> -2\sqrt{\lambda_1\lambda_2}.
\label{eq:constraints1}
\end{split}
\end{align}
There is also the condition that the squared masses should be positive, to ensure that the extremum of the potential is actually a minimum. This gives the additional constraints\ \cite{Nie:1998yn}:
\begin{align}
\begin{split}
\lambda_3 &< 0, \\
\lambda_4 &> \lambda_3, \\
\lambda_3+\lambda_5 &< 2\sqrt{\lambda_1\lambda_2}.
\label{eq:constraints2}
\end{split}
\end{align}

Having defined the model, we can look at the amount of fine-tuning present in the theory according to the fine-tuning measures discussed above.

\subsection{Evaluation of the Dekens Measure}
In this section we will first determine what the proper way is to use the Dekens measure and conclude how much fine-tuning is found in this way. As mentioned in Section \ref{sec:measures}, when using the Dekens measure we have to split the parameters of the model in two sets. There is no prescription for how we should make this splitting. Therefore we will look at two different cases. In both cases we will solve the minimum equations for the two $\mu^2$ parameters, but then we make different choices for the $q_i$. Due to the fact that there are two minimum equations, one has to select two dependent parameters $q_i$. 

\textbf{Case I: $q_i = \{\mu_1$, $\mu_2\}$}\\
When we choose $\mu_1$ and $\mu_2$ as the dependent variables, we can derive the following expressions for the Dekens measure:
\begin{align}
\begin{split}
\Delta_D^I(\mu_1,\lambda_1) &= \frac{1}{2}\frac{\lambda_1v_1^2}{\lambda_1 v_1^2 + \lambda_+ v_2^2}, \\
\Delta_D^I(\mu_1,\lambda_3) &= \frac{1}{2}\frac{\lambda_3v_2^2}{\lambda_1 v_1^2 + \lambda_+ v_2^2}, \\
\Delta_D^I(\mu_1,\lambda_5) &= \frac{1}{2}\frac{\lambda_5v_2^2}{\lambda_1 v_1^2 + \lambda_+ v_2^2}, \\
\Delta_D^I(\mu_1,v_1) &= \frac{\lambda_1v_1^2}{\lambda_1 v_1^2 + \lambda_+ v_2^2}, \\
\Delta_D^I(\mu_1,v_2) &= \frac{\lambda_+v_2^2}{\lambda_1 v_1^2 + \lambda_+ v_2^2}, \\
\end{split}
\quad\quad\quad
\begin{split}
\Delta_D^I(\mu_2,\lambda_2) &= \frac{1}{2}\frac{\lambda_2v_2^2}{\lambda_2 v_2^2 + \lambda_+ v_1^2}, \\
\Delta_D^I(\mu_2,\lambda_3) &= \frac{1}{2}\frac{\lambda_3v_1^2}{\lambda_2 v_2^2 + \lambda_+ v_1^2}, \\
\Delta_D^I(\mu_2,\lambda_5) &= \frac{1}{2}\frac{\lambda_5v_1^2}{\lambda_2 v_2^2 + \lambda_+ v_1^2}, \\
\Delta_D^I(\mu_2,v_1) &= \frac{\lambda_2v_2^2}{\lambda_2 v_2^2 + \lambda_+ v_1^2}, \\
\Delta_D^I(\mu_2,v_2) &= \frac{\lambda_+v_1^2}{\lambda_2 v_2^2 + \lambda_+ v_1^2}.\\
\end{split}
\end{align}
The final amount of fine-tuning is defined as: $\Delta^I_D = \underset{i,j}{\max} \ |\Delta^I_D(\mu_i,p_j)|$.

In all cases the measure produces a ratio, where one of the contributions to the $\mu$ parameter is compared to the actual value of the $\mu$ parameter. If a contribution is much larger than the value itself, there has to be a large cancellation between terms, which would correspond to fine-tuning.

Since there is always one large and one small contribution to $\mu_i$, we can discard most of the terms already. In addition, some terms are simply multiples of another. Therefore the relevant terms are:
\begin{equation}
\Delta^I_D = \max\left\{\left|\frac{\lambda_1v_1^2}{\lambda_1 v_1^2 + \lambda_+ v_2^2}\right|, \left| \frac{1}{2}\frac{\lambda_3v_1^2}{\lambda_2 v_2^2 + \lambda_+ v_1^2} \right|, \left| \frac{1}{2}\frac{\lambda_5v_1^2}{\lambda_2 v_2^2 + \lambda_+ v_1^2} \right|, \left| \frac{\lambda_+v_1^2}{\lambda_2 v_2^2 + \lambda_+ v_1^2} \right|\right\}.
\end{equation}
But all of these ratios will be \order{1}, since the $v_1$ terms dominates in all the denominators. The only way to get a number much larger than one is by having a large ratio $\lambda_3/\lambda_+$ or $\lambda_5/\lambda_+$. In general these ratios will not be very large, so we do not expect any fine-tuning. The case where these fractions are large will be discussed separately later.

\textbf{Case II: $q_i = \{\lambda_1$, $\lambda_2\}$}\\
Now we take a look at a case where we calculate fine-tuning with respect to a different set of parameters than the ones we used to solve the minimum equations. First we rewrite the minimum equations to get expressions for this set of parameters:
\begin{align}
\begin{split}
\lambda_1 = \frac{\mu_1^2-\lambda_+ v_2^2}{v_1^2}, \\
\lambda_2 = \frac{\mu_2^2-\lambda_+ v_1^2}{v_2^2}. \\
\end{split}
\end{align}
Calculating the different expressions for the fine-tuning measure gives:
\begin{align}
\begin{split}
\Delta_D^{II}(\lambda_1,\lambda_3) &= -\frac{\lambda_3v_2^2}{2(\mu_1^2-\lambda_+v_2^2)} = \frac{\lambda_3v_2^2}{2\lambda_1v_1^2}, \\
\Delta_D^{II}(\lambda_1,\lambda_5) &= -\frac{\lambda_5v_2^2}{2(\mu_1^2-\lambda_+v_2^2)} = \frac{\lambda_5v_2^2}{2\lambda_1v_1^2}, \\
\Delta_D^{II}(\lambda_1,v_1) &= -2, \\
\Delta_D^{II}(\lambda_1,v_2) &= 2\frac{\lambda_+v_2^2}{\mu_1^2-\lambda_+v_2^2} = \frac{2\lambda_+v_2^2}{\lambda_1v_1^2}, \\
\Delta_D^{II}(\lambda_1,\mu_1) &= \frac{2\mu_1^2}{\mu_1^2-\lambda_+v_2^2} = \frac{2(\lambda_1v_1^2 + \lambda_+v_2^2)}{\lambda_1v_1^2}, 
\end{split}
\quad\quad
\begin{split}
\Delta_D^{II}(\lambda_2,\lambda_3) &= -\frac{\lambda_3v_1^2}{2(\mu_2^2-\lambda_+v_1^2)} = \frac{\lambda_3v_1^2}{2\lambda_2v_2^2}, \\
\Delta_D^{II}(\lambda_2,\lambda_5) &= -\frac{\lambda_5v_1^2}{2(\mu_2^2-\lambda_+v_1^2)} = \frac{\lambda_5v_1^2}{2\lambda_2v_2^2}, \\
\Delta_D^{II}(\lambda_2,v_1) &= 2\frac{\lambda_+v_1^2}{\mu_2^2-\lambda_+v_1^2} = \frac{2\lambda_+v_1^2}{\lambda_2v_2^2}, \\
\Delta_D^{II}(\lambda_2,v_2) &= -2, \\
\Delta_D^{II}(\lambda_2,\mu_2) &= \frac{2\mu_2^2}{\mu_2^2-\lambda_+v_1^2} = \frac{2(\lambda_2v_2^2 + \lambda_+v_1^2)}{\lambda_2v_2^2}, 
\end{split}
\raisetag{5.5\normalbaselineskip}
\end{align}
where we have written all expressions in terms of the independent parameters in the second step. The total amount of fine-tuning is again defined as the maximum of the absolute value of all the terms: $\Delta_D^{II} = \underset{i,j}{\max}\ |\Delta_D^{II}(\lambda_i,p_j)|$. As some of the expressions with $\lambda_2$ depend on the large ratio $v_1^2/v_2^2$, this will result in a large value for the fine-tuning measure.

So we see that when analyzing the model in two different ways, two completely different results for the amount of fine-tuning in the model can be obtained. This then raises the questions: is one of these ways to apply the Dekens measure more appropriate than the other? Or does one have to check all cases and find the maximum, as done in\ \cite{Dekens:2014ina}? To investigate this, we will turn to a simplified case where we can easily compare all the different scenarios.

\subsubsection{Analysis of the origin of the fine-tuning}
We will start by looking at the minimum equation for $\mu_2^2$, since this turns out to be the source of the discrepancy. Dividing both sides by $v_1^2$ yields:
\begin{equation}
\frac{\mu_2^2}{v_1^2} = \frac{v_2^2}{v_1^2}\lambda_2 + \lambda_+. \label{eq:2HDM_toymodel}
\end{equation}
Then we will rename the different terms in this equation for clarity. We will replace all \order{1} parameters by capital letters, and denote the small fraction $v_2^2/v_1^2$ by $x$, so the minimum equation is now written schematically as: $A = xB + C$. 

When solving the minimum equation for $A$, we see that one can choose $B$ and $C$ to be numbers of \order{1}, such that $A$ will also be \order{1}. When taking $A$ as the dependent parameter, and $B$, $C$ and $x$ as the independent parameters, we obtain for the Dekens measure:  
\begin{align}
\Delta_D(A) = \max\left\{ \left| \frac{xB}{xB+C} \right|, \left| \frac{C}{xB+C} \right|, \left| \frac{xB}{xB+C}\right| \right\} = \max\{ \mathcal{O}(x), \mathcal{O}(1), \mathcal{O}(x) \} = \mathcal{O}(1).
\end{align}
This is completely analogous to Case I in the 2HDM discussion.

Now we will look at the scenario similar to Case II. We again solve the equation with respect to $A$, but then take $B$ as the dependent parameter: 
\begin{equation}
B = \frac{A-C}{x}.
\end{equation}
Now we find the fine-tuning to be:
\begin{equation}
\Delta_D(B) = \max\left\{ \left| \frac{xA}{A-C}\frac{1}{x} \right|, \left| \frac{xC}{A-C}\frac{1}{x} \right|, \left| \frac{x}{B}\frac{A-C}{x^2}\right| \right\} = \max\left\{ \left| \frac{A}{xB} \right|, \left| \frac{C}{xB} \right|, 1 \right\} = \mathcal{O}(1/x).
\end{equation}
So in this case we get a large amount of fine-tuning just like in Case II\footnote{In both scenarios the fine-tuning is calculated with respect to $x$. Writing $x$ as $v_2^2/v_1^2$ and calculating the fine-tuning with respect to the two vevs separately gives the same result.}. But is this large value for the Dekens measure really due to fundamental fine-tuning in the parameters of the theory?

To understand this result, we look at two ways to interpret this situation. We will first compare the size of the terms on the left-hand side of the equation to the right-hand side, and then look at the effect of variations of parameters.

When comparing the contributions on the left-hand side of the equation to the right-hand side, we see that in the original equation $A = xB + C$, we have two contributions to $A$: a small one ($xB$) and a large one ($C$). So $A$ will be almost exactly the same as $C$, with only a minor correction due to $xB$. There is no large cancellation, so no large fine-tuning. But in Case II we set $B$ as the parameter of interest, and then we see that there is a huge cancellation between $A$ and $C$ in order to get the \order{1} value for $B$. However this is not due to fine-tuning, it is due to rewriting an equation where the $B$ term was simply a small correction. So we see that there are multiple ways to use the minimum equations to study the dependence of parameters on other parameters, but we argue that if there is a way to do this without large fine-tuning, then it means that there simply is no large fine-tuning. Other ways may lead to a large Dekens measure, but that is just a consequence of the choice of $q_i$ parameters and considered only apparent fine-tuning.

We now look at variations in the parameters. If a variation of \order{1} in $A$ can only be achieved by varying $B$ then a large variation in $B$ would be necessary, indicating a large amount of fine-tuning. However, the variation in $A$ can also be achieved by varying $C$, where only an \order{1} change would be necessary. To demand that the variation in $A$ of \order{1} has to come from a variation only in $B$, is simply an artificial requirement. Again we see apparent fine-tuning appearing, but this does not mean there is actual fine-tuning in the theory.  

Therefore, our conclusion on applying the Dekens measure is that one should use as dependent parameters the same parameters that were used to solve the minimum equations for (Case I in our discussion), otherwise there is the risk of finding apparent fine-tuning.

\subsection{The Barbieri-Giudice measure}
We will evaluate the BG measure using two observables: the masses of the two CP-even scalar bosons $h$ and $H$. These are the only two interesting masses, since we found that the other masses depend on the input parameters in such a way that no fine-tuning is possible.

\begin{figure}
    \centering
    \begin{subfigure}[b]{0.48\textwidth}
        \includegraphics[width=\textwidth]{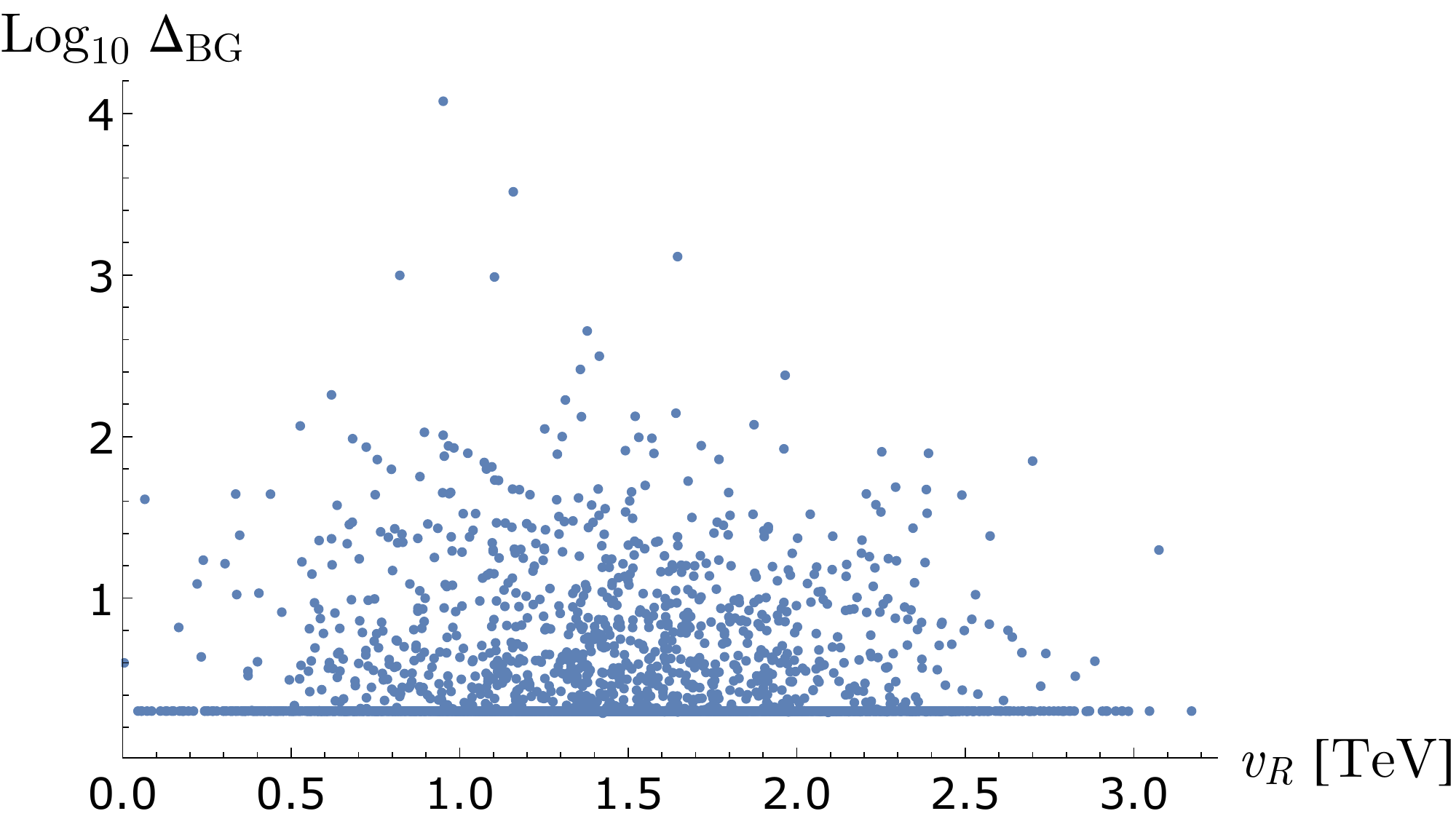}
        \caption{}
        \label{fig:lightHiggs_2HDM_scatter}
    \end{subfigure}
  \quad
    \begin{subfigure}[b]{0.48\textwidth}
        \includegraphics[width=\textwidth]{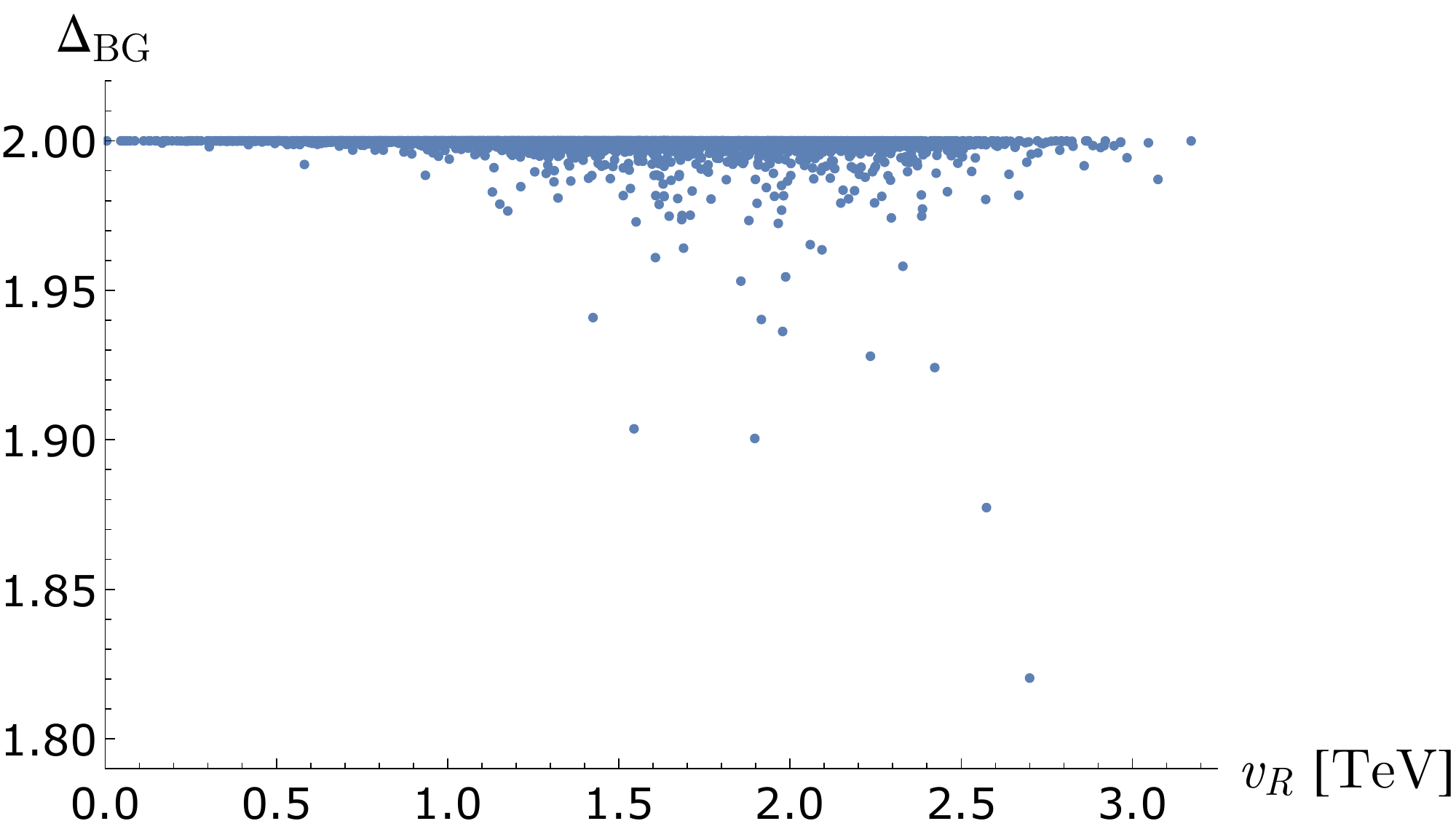}
        \caption{}
        \label{fig:heavyHiggs_2HDM_scatter}
    \end{subfigure}
    \caption{Numerical result for the BG measure using the mass of the neutral CP-even Higgs bosons as observables. Figure (a) and (b) show the result for the light Higgs boson $h$ and the heavier $H$ respectively, as a function of $v_2$. In these figures we have set $v_1^2 + v_2^2 = 50$ TeV. Note that Figure (a) uses a logarithmic scale for $\Delta_{BG}$, whereas Figure (b) uses a linear scale.} \label{fig:BGHiggs_2HDM}
\end{figure}

The results of evaluating the BG measure with these two masses as observables are  shown in Figure \ref{fig:BGHiggs_2HDM}. We see here that for the lightest Higgs boson most points have a negligible amount of fine-tuning of $\Delta_{BG} \approx 2$, but there are also some points with a large amount of fine-tuning. We can understand this behavior from the approximate expression for the Higgs mass:
\begin{equation}
	m_h^2 \approx \left[2\lambda_2 -\frac{(\lambda_3+\lambda_5)^2}{2\lambda_1}\right]v_2^2.
\end{equation}
First of all we note that this mass is not sensitive to the large scale $v_1$, so the fine-tuning we see is not due to the large hierarchy in the theory. However, there is another source of fine-tuning present. If there is a large cancellation in the term $2\lambda_2 - \frac{(\lambda_3+\lambda_5)^2}{2\lambda_1}$,  we will have $m_h^2 \ll v_2^2$, so the individual contributions are much larger than the final value. 

For the heavy boson $H$ we see a hard cap at $\Delta_{BG} = 2$. We can again understand this from the approximate expression for the mass:
\begin{equation}
	m_H^2 \approx 2\lambda_1v_1^2 +\frac{(\lambda_3+\lambda_5)^2}{2\lambda_1}v_2^2
\end{equation}
All contributions to this mass are positive, so there is no way to get a cancellation. Because of the quadratic dependence on $v_1$, the logarithmic derivative in the BG measure produces a factor of 2 which leads to the hard cutoff seen in Figure \ref{fig:heavyHiggs_2HDM_scatter}. Actually all points have a value of $\Delta_{BG}$ slightly smaller than 2, since the BG measure compares the terms on the right-hand side to the actual value of $m_H^2$. Due to the presence of the small $v_2^2$ contribution, $m_H^2$ will be slightly larger than $2\lambda_1 v_1^2$, resulting in $\Delta_{BG} < 2$.

\subsection{The case of large Dekens measure}
\label{sec:LargeDekens2HDM}
So far we have seen a model where the Dekens measure shows no signs of fine-tuning, but the BG measure shows that in some cases there can be fine-tuning. One could therefore argue that it might be enough to just use the BG measure, since the Dekens measure does not show any fine-tuning. But we will now highlight a case where the Dekens measure does produce a large value.

We can accomplish this by having large ratios $\lambda_3/\lambda_+$ and $\lambda_5/\lambda_+$. This is the case when $\lambda_5 \approx -\lambda_3$. This is allowed, since $\lambda_3$ has to be negative, while $\lambda_5$ can be positive. When these two terms cancel to a large degree, $\mu_2^2$ will no longer be of \order{v_1^2}, but can be as small as \order{v_2^2}. This will result in a large value for the Dekens measure in Case I, since there is a large cancellation between two terms of \order{v_1^2} that results in a much smaller value.

The question is now whether this fine-tuning also shows up in the BG measure. It turns out that the BG measure is not sensitive to this fine-tuning. We already concluded that the fine-tuning in the BG measure is due to a cancellation between a combination of $\lambda$'s: $2\lambda_2 -\frac{\lambda_+^2}{2\lambda_1} \approx 0$. But now we have the situation where $\lambda_+ \approx 0$. So with $\lambda_2 \sim$ \order{1}, there will never be such a cancellation and the BG measure will not find any fine-tuning. This was confirmed numerically. This means that both measures are necessary in order to capture different sources of fine-tuning in the theory. 

\subsection{Conclusions on fine-tuning in the 2HDM}
Our investigation of the 2HDM has shown that one has to be very careful when determining the amount of fine-tuning in the minimum equations. A large hierarchy between parameters does not automatically mean that there is fine-tuning present. Actually, in the 2HDM it is possible to solve the minimum equations in a way such that all coupling constants are natural and perturbative, but without introducing any fine-tuning in the minimum equations.

Interestingly, when evaluating the BG measure, we see that the masses of the scalar bosons are not sensitive to the hierarchy in scales, but there are other sources of fine-tuning that are not visible when using the Dekens measure. On the other hand, there are situations possible where the Dekens measure becomes large for natural and perturbative parameters due to large cancellations, while this is absent in the BG measure. This shows that both measures need to be considered in order to be sure about the absence of fine-tuning in the theory, as they can capture different sources of fine-tuning. 

\section{Fine-tuning in the left-right symmetric model}
Now that we understand how fine-tuning works in the 2HDM, we can apply those lessons to a model that is more complex. To that end, we will now look at the (parity  conserving) left-right symmetric model (LRSM) \cite{Pati:1974yy,Mohapatra:1974gc}. A previous discussion of fine-tuning in this model claimed that there was fine-tuning as high as $\Delta \sim 10^{20}$ in this model \cite{Dekens:2014ina}, while other papers claim a fine-tuning of at least $\Delta \sim 10^7$ \cite{Deshpande:1990ip}. We now want to investigate this as we did for the 2HDM.
Since we are only interested in the Higgs sector of the model, we will not discuss the full LRSM, but only the relevant parts. For a full review see e.g.\ \cite{Dekens:2014ina,Senjanovic:1978ev}.

\subsection{The LRSM Higgs sector}
The gauge group of the Standard Model is $SU(3)_C\times SU(2)_L\times U(1)_Y$. In the LRSM, this gauge group is extended to the gauge group $SU(3)_c\times SU(2)_L\times SU(2)_R\times U(1)_{B-L}$  at high energies. Since we are only interested in the electroweak sector, the $SU(3)_C$ group will not be considered here. The representations will therefore be given in terms of the gauge group $SU(2)_L\times SU(2)_R\times U(1)_{B-L}$. 

The Higgs doublet in the Standard Model is now replaced by a bidoublet $\phi$ in the $(2,2^*,0)$ representation. Two additional scalar fields are added to break the LR gauge group to the SM gauge group: $\Delta_L \in (3,1,2)$ and $\Delta_R \in (1,3,2)$. We can write these fields in terms of complex scalars:
\begin{equation}
\phi = \begin{pmatrix}
\phi_1^0 & \phi_1^+ \\
\phi_2^- & \phi_2^0
\end{pmatrix},
\qquad
\Delta_{L,R} = \begin{pmatrix}
\delta_{L,R}^+/\sqrt{2} & \delta_{L,R}^{++} \\
\delta_{L,R}^0 & -\delta_{L,R}^+/\sqrt{2}
\end{pmatrix}.
\end{equation}
From these fields we can construct a potential that is invariant under the gauge group. In addition to this invariance, we also demand invariance under $P$-symmetry, resulting in the following potential:
\begin{align}
\begin{split}
V_H^P = &-\mu_1^2\Aone - \mu_2^2 \left[\Atwo + \Athree \right] -\mu_3^2\left[\AL + \AR\right] \\
&+ \lambda_1 \left[\Aone\right]^2 + \lambda_2 \left( \left[\Atwo\right]^2 + \left[\Athree\right]^2 \right) + \lambda_3\Atwo\Athree \\
&+ \lambda_4\Aone\left[ \Atwo + \Athree \right] + \rho_1\left( \left[\AL\right]^2 + \left[\AR\right]^2 \right)\\ 
&+ \rho_2\left[ \BLL + \BRR \right] + \rho_3\AL\AR \\
&+ \rho_4\left[\BLR + \BRL\right] \\
&+ \alpha_1\Aone\left[\AL + \AR\right] \\
&+ \alpha_2\left( e^{i\delta_2}\left[\Atwo\AR + \Athree\AL\right]+ h.c. \right) \\
&+ \alpha_3 \left[\CL + \CR\right] + \beta_1\left[\Done + \Dones\right] \\
&+ \beta_2\left[\Dthree + \Dthrees\right] + \beta_3\left[\Dtwo + \Dtwos\right]
\end{split}
\label{eq:LRpotential}
\end{align}
The vacuum expectation values of the fields are defined as:
\begin{equation}
\langle\phi\rangle = \frac{1}{\sqrt{2}}\begin{pmatrix}
	\kappa & 0 \\
	0 & \kappa'e^{i\alpha}
\end{pmatrix},
\qquad
\langle\Delta_L\rangle = \frac{1}{\sqrt{2}}\begin{pmatrix}
	0 & 0 \\
	v_Le^{i\theta_L} & 0
\end{pmatrix},
\qquad
\langle\Delta_R\rangle = \frac{1}{\sqrt{2}}\begin{pmatrix}
	0 & 0 \\
	v_R & 0
\end{pmatrix},
\end{equation}
where the two phases can be sources of spontaneous CP violation. We will set these phases to zero to simplify the discussion. 

In this model, there is a clear hierarchy in the scales: $v_R$ is the highest scale in the theory, since it is responsible for the breaking of the $SU(2)_R \times U(1)_{B-L}$ symmetry. It needs to have a value of at least a few TeV. The second to highest scales in the theory are $\kappa$ and $\kappa'$. They are responsible for electroweak symmetry breaking, so they must be \order{100 \text{ GeV}}. The value of $v_L$ is related to the Majorana mass of the neutrinos. In addition to this type-II seesaw mechanism, there is also a type-I seesaw, giving mass through the addition of heavy right-handed neutrinos. Since we want to avoid any additional fine-tuning in the model, we will assume that these two contributions do not give rise to large cancellations. Therefore, we will assume $v_L$ to be ${\cal O}(1 \text{ eV})$.

The minimum equations for this potential are given by:
\begin{align}
\begin{split}
	\frac{\mu_1^2}{v_R^2} &= \frac{\alpha_1}{2} - \frac{\kappa'^2}{2(\kappa^2-\kappa'^2)}\alpha_3 + \frac{\kappa^2+\kappa'^2}{v_R^2}\lambda_1 + 2\frac{\kappa\kappa'}{v_R^2}\lambda_4, \\
	\frac{\mu_2^2}{v_R^2} &= \frac{\alpha_2}{2} + \frac{\kappa\kappa'}{4(\kappa^2+\kappa'^2)}\alpha_3 + \frac{\kappa^2+\kappa'^2}{2v_R^2}\lambda_4 + \frac{\kappa\kappa'}{v_R^2}(\lambda_3+2\lambda_2), \\
	\frac{\mu_3^2}{v_R^2} &=	\rho_1 + \frac{\kappa^2+\kappa'^2}{2v_R^2}\alpha_1 + \frac{2\kappa\kappa'}{v_R^2}\alpha_2 + \frac{\kappa'^2}{2v_R^2}\alpha_3, \\
	2\rho_1 - \rho_3 &= \frac{1}{v_Rv_L}(\kappa\kappa'\beta_1 + \kappa^2\beta_2 + \kappa'^2\beta_3),
\end{split}
\label{eq:LRminEqn}
\end{align}
where we should note that the first three equations are approximate expressions in which terms of order $v_L/v_R$ and higher were neglected. This will not influence our conclusions about the fine-tuning in this model.

We need to ensure again that the potential is bounded from below. In general, finding the constraints on the coupling constants to ensure boundedness is a difficult problem. Therefore we will use a different approach than in the 2HDM to ensure boundedness. Since we impose the values of the vevs, our system of minimum equations is a linear system in the parameters of the potential. Thus it has a unique solution, which means that there cannot be multiple extrema of the potential. Therefore we can ensure boundedness by making sure that the extremum we find is a minimum. We do this by investigating the scalar masses. When all masses are positive, we are at a minimum of the potential, and we know that the potential will be bounded from below. The masses will be evaluated numerically, since it is intractable to obtain analytic expressions for the masses even in the above approximation of dropping terms of order $v_L/v_R$ and higher. 

\subsection{Evaluation of the Dekens measure}
We will start our fine-tuning discussion by looking at the Dekens measure in this model. We want to find out if it is possible in this model to have  natural and perturbative coupling constants, to have the hierarchy in the vevs as described above, but without having fine-tuning in the theory.

We can apply most of our conclusions from the 2HDM, at least for the first three minimum equations. If we look at these three equations, we see that we are in nearly the same situation as in the 2HDM. In all of these equations, there is at least one term of \order{1}, and the other terms are much smaller than 1. Therefore, all $\mu^2$ parameters will be \order{v_R^2}, so they will have a value similar to the highest scale in the theory. Since in general there are no cancellations appearing in these equations, we again expect to see no fine-tuning.

But the last equation has a different structure than what we have seen so far. There is no mass parameter in this equation, so it is not so clear for which parameter we should solve this equation. Furthermore, there is a clear hierarchy present in this equation.

If we solve this minimum equation for $\rho_1$, we obtain:
\begin{equation}
\rho_1 = \frac{1}{2v_Lv_R}(\kappa\kappa'\beta_1 + \kappa^2\beta_2 + \kappa'^2\beta_3) + \frac{\rho_3}{2}.
\end{equation}
By taking values of \order{1} for the $\beta$ parameters, and taking $v_R$ to be \order{10 \text{ TeV}}, we will get a value for $\rho_3$ of $\mathcal{O}(\kappa^2/(v_L v_R))$ = \order{10^9}.  This is in conflict with our demand that all coupling constants should be \order{1}. The only way to avoid this non-perturbative value for $\rho_1$ is to fine-tune the values of the $\beta$'s in such a way that they cancel up to 9 significant digits. When we apply the Dekens measure as argued before, this would give a large value for the Dekens measure, since the individual $\beta$ parameters would have a much larger contribution to $\rho_1$ than the actual value of $\rho_1$. This is the argument for claiming that there is a large amount of fine-tuning in the LRSM (see e.g. \cite{Dekens:2014ina, Deshpande:1990ip, Umezawa:1997yi, Barenboim:1998ib}).

However, this is not the only way we can solve this equation. Let us see what happens when we solve for one of the $\beta$ parameters:
\begin{equation}
\beta_1 = -\frac{\kappa}{\kappa'}\beta_2 - \frac{\kappa'}{\kappa}\beta_3 + \frac{v_Lv_R}{\kappa\kappa'}(2\rho_1 -  \rho_3).
\label{eq:seesaw_beta}
\end{equation}
Now we can take all coupling constants on the right-hand side of the equation \order{1}, and we will find a value for $\beta_1$ which is also \order{1}. The two $\rho$ parameters are simply small contributions that only contribute at the $10^{-9}$ level.

In this situation there is no fine-tuning. We see here that solving the minimum equations in a different way can lead to very different conclusions. We argue that a model has no fine-tuning if there is a way to solve the minimum equations such that all coupling constants are \order{1} without introducing fine-tuning. So in this case if we solve for one of the $\beta$ parameters we can have all coupling constants perturbative and natural without fine-tuning, so we conclude that if the Dekens measure is used in the proper way, one does not find fine-tuning in the LRSM. In practice, one could simply study different cases and if one or more yield $\Delta$ of \order{1}, then it means there is a way to solve the problem without fine-tuning for natural and perturbative coupling constants.  

It is important to emphasize that rewriting this minimum equation is allowed because it is a constraint equation. Such an equation is used to eliminate a parameter from the system, but there is no prescription for which parameter to eliminate. In principle, every way to solve a minimum equation is equally fine, but for some choices it seems like there is fine-tuning, while for other choices this is not the case. If there is a way to obtain a Dekens measure of \order{1}, then that means there is a way of satisfying the minimum equations without fine-tuning, i.e.\ with \order{1} parameters without large cancellations and we argue that that means that the minimum equations are then not a source of fine-tuning.

\subsection{Evaluation of the Barbieri-Giudice Measure}
Just like in the 2HDM, we can see how the results of the Dekens measure compare to the BG measure. The difference is that in this case it is not possible to get analytic expressions for the masses. We can only get numerical results for the masses. Also the derivatives in the BG measure have to be evaluated numerically. 

The BG measure will now be computed as the maximum of the values for the individual masses:
\begin{equation}
\Delta_{BG} = \max_{i,j}\left\lvert\frac{p_j}{m_i^2}\frac{\partial m_i^2}{\partial p_j}\right\rvert.
\end{equation} 
Whereas there were only two masses with non-trivial expressions in the 2HDM case, now we have 14 masses that all depend non-trivially on the input parameters. Because of this, we expect that the amount of fine-tuning spreads over a larger range. This is because the chance is now higher that there is some cancellation in one of the masses. This is exactly the behavior we see in Figure \ref{fig:BG_LRSM_NoConstraint}. One should also keep in mind that the more points are sampled, the higher the chance to find some case that happens to show some cancellation and hence fine-tuning. From the figure it is clear though that the generic model will have a fine-tuning below 100. 

In this analysis the dimensionless parameters have been sampled from a combined distribution consisting of two log-normal distributions, one centered around 1 and the other centered around -1. Constraints were imposed on some parameters to ensure boundedness of the potential. 

\subsubsection{Imposing the Higgs mass constraint}
\begin{figure}
    \centering
    \begin{subfigure}[b]{0.49\textwidth}
        \includegraphics[width=\textwidth]{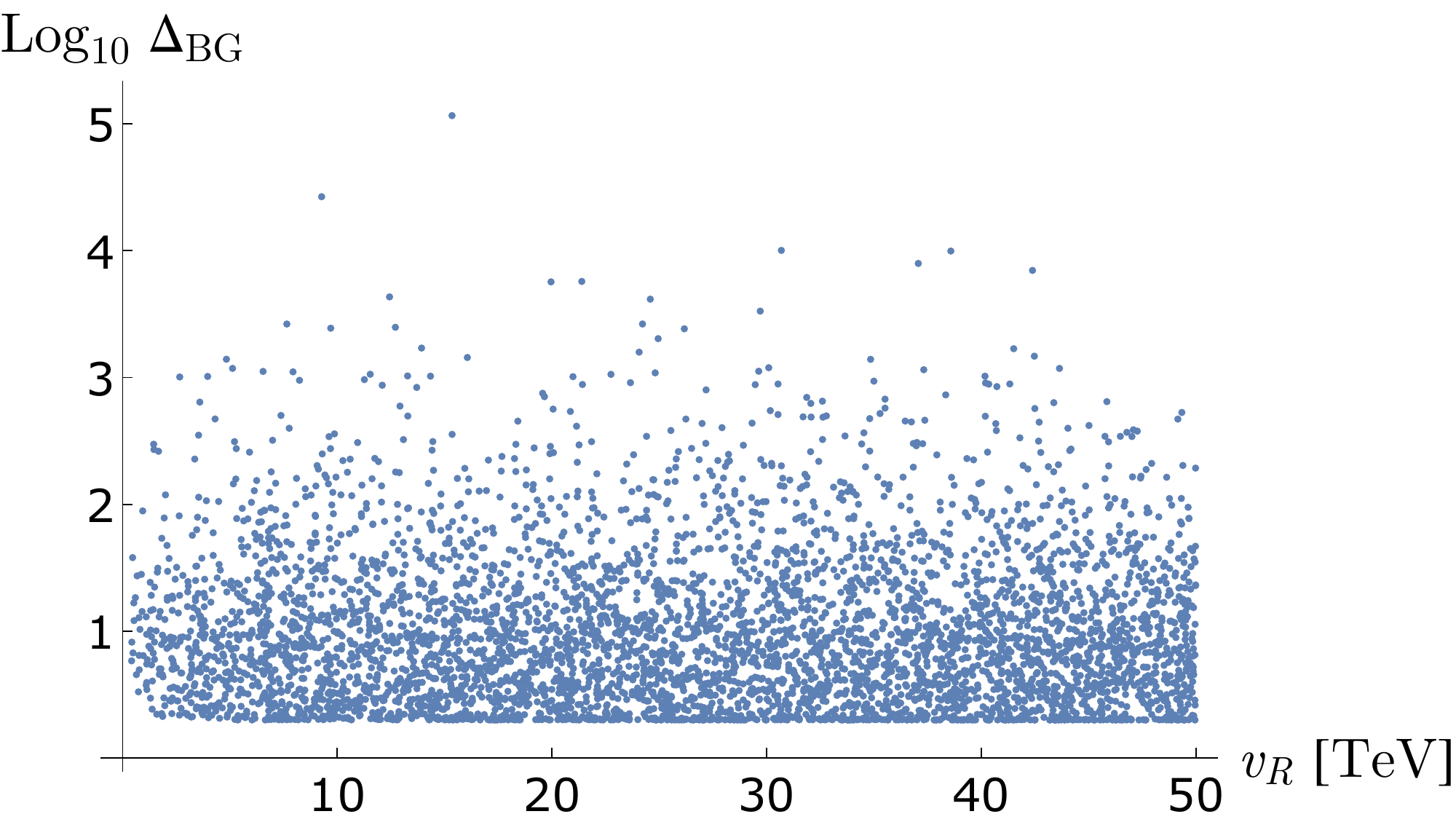}
        \caption{}
        \label{fig:BG_LRSM_NoConstraint}
    \end{subfigure}
  \vspace{5pt}
	\begin{subfigure}[b]{0.49\textwidth}
        \includegraphics[width=\textwidth]{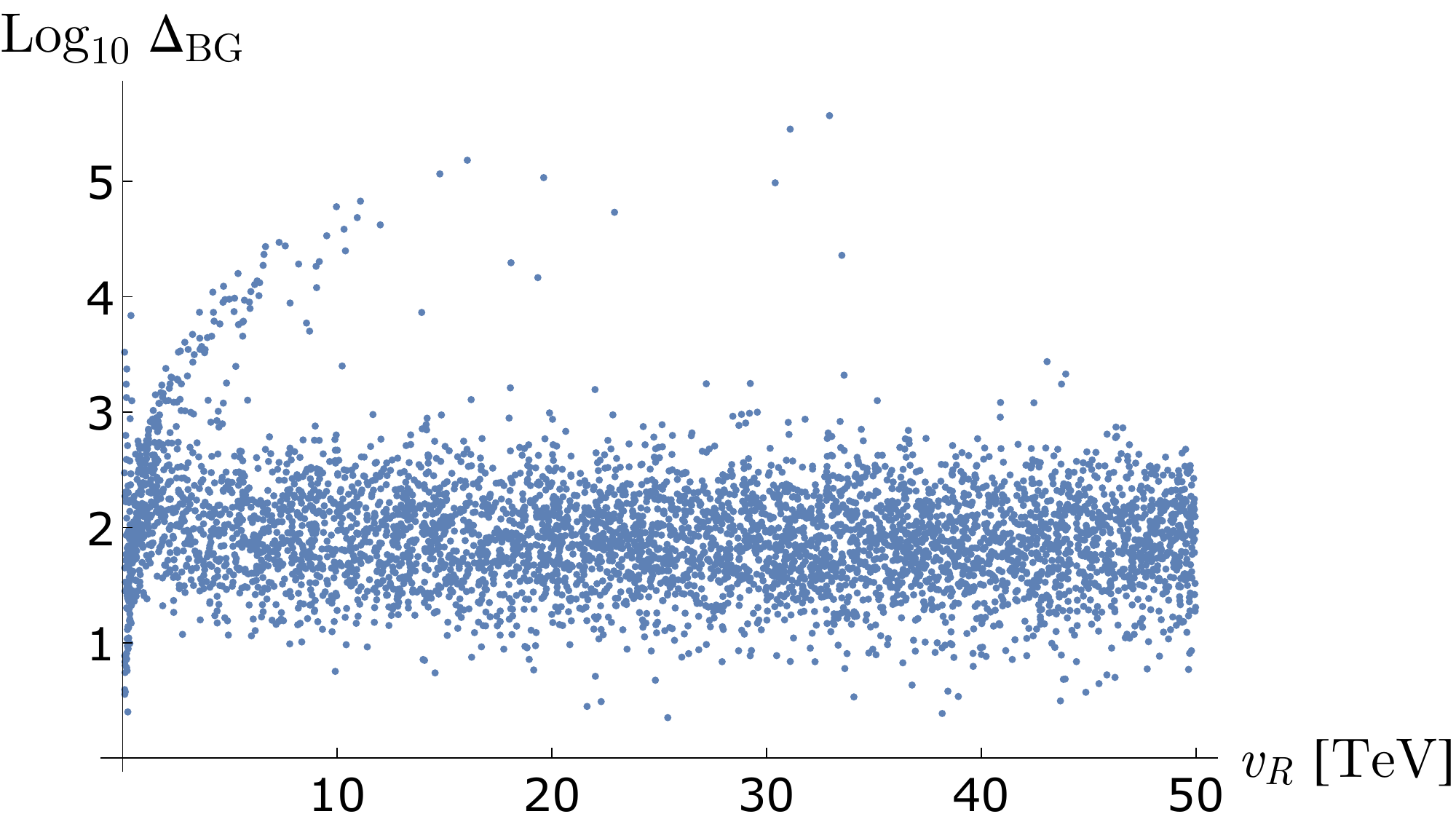}
        \caption{}
        \label{fig:BG_LRSM_Constraint}
    \end{subfigure}     
    \caption{Numerical result of evaluating the BG measure of 5000 points in the LRSM with (a) no constraint imposed on the Higgs mass and (b) the constraint $100 \text{ GeV}< m_h < 150 \text{ GeV}$.}\label{fig:BG_LRSM}
\end{figure}

In order for the LRSM model to be phenomenologically viable, we wish to impose that the lightest Higgs mass is close to 125 GeV\ \cite{Chatrchyan:2012xdj,Aad:2012tfa}. Now we will check if this constraint has an impact on the amount of fine-tuning. Because the computation of a single model point takes a relatively long time, we will not impose a very strict constraint on the Higgs mass. We require that the mass is between 100 GeV and 150 GeV. The result of calculating the BG measure in model points with this constraint is shown in Figure \ref{fig:BG_LRSM_Constraint}.

There are two observations we can make based on this figure. First of all, we see that most points have a larger amount of fine-tuning than in the case without constraint on the Higgs mass. Even though we do not have analytic expressions for the masses, we can still make observations on the cause for this increase in fine-tuning. This moderate amount of fine-tuning is not due to the hierarchy between $\kappa, \kappa'$ and $v_R$, since increasing the value of $v_R$ has no effect on the mass of the lightest scalar boson in the bulk of the points. Our explanation for this increase in fine-tuning is that the mass of the lightest Higgs boson is a sum of multiple terms, all of which are \order{\kappa^2}. If there are a lot of these terms, the sum can be significantly larger than $\kappa^2$, such that some cancellations are necessary in order to get a Higgs mass in the correct range.

The other observation is that while the bulk of the points are uniformly distributed across the $v_R$ range, there is another set of points showing a clear line in the $(\Delta_{BG},v_R)$-plane. These points are focused around small values for $v_R$. When we investigate these points, we see that they have small values for both $v_R$ and the combination $2\rho_1-\rho_3$. In these cases, there are contributions to the mass of the lightest Higgs boson of the order $(2\rho_1-\rho_3)v_R^2$. When this term becomes small it starts interfering with the other contributions to $m_h^2$, that are of \order{\kappa^2}. There needs to be some cancellation between the $\rho$ terms in order to get a correct Higgs mass for these points. For larger values of $v_R$, the fine-tuning between the two $\rho$ terms needs to increase, leading to the higher values for $\Delta_{BG}$ and the lower density of points in the plot.

\subsection{Large Dekens measure}
Just like in the 2HDM it is possible to get a large value for the Dekens measure, while keeping the BG measure low. We will consider two distinct ways to achieve a large Dekens measure. First we will see what happens when we artificially introduce a large Dekens measure in one of the equations for a $\mu^2$ parameter, which is very similar to the 2HDM case. Then we will investigate the case where there is a large cancellation in the seesaw relation.

\begin{figure}
   \centering
        \includegraphics[width=0.6\textwidth]{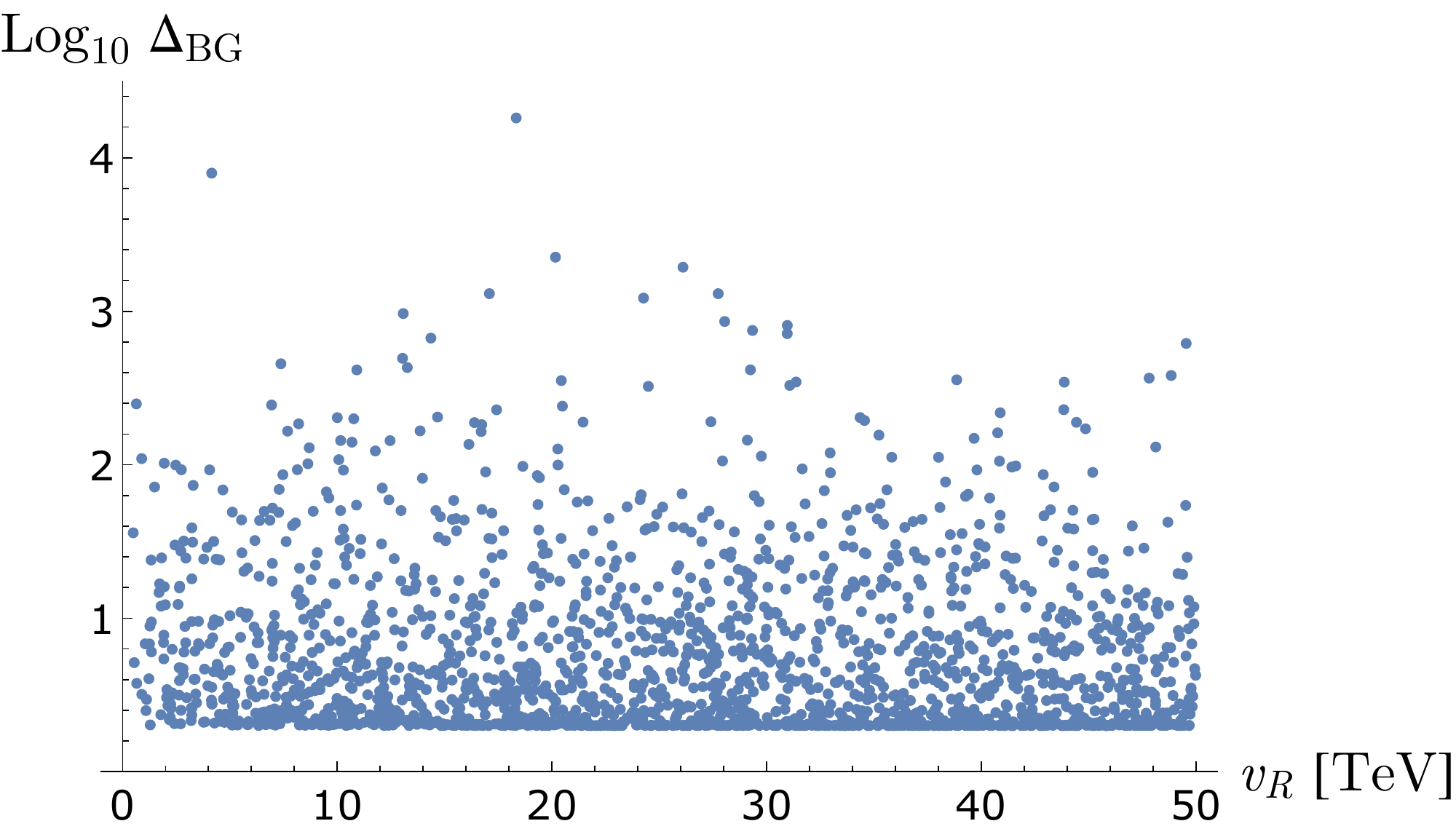}
  \caption{Numerical result of evaluating the BG measure of 2000 points in the LRSM, where a cancellation has been imposed in the formula for $\mu_1^2$.} \label{fig:MuCancellation_LRSM}
    \label{fig:BGHighDekens_LRSM}
\end{figure}

If we look at the equation for $\mu_1^2$ (Eq.\ (\ref{eq:LRminEqn})), we see that it is possible to have a large cancellation while keeping all coupling constants \order{1}. We can do this by having a value of $\alpha_3$ such that it cancels the $\alpha_1$ contribution. Then we add a small term to $\alpha_3$ in order to not have perfect cancellation with $\alpha_1$:
\begin{equation}
\alpha_3 = \frac{\kappa^2-\kappa'^2}{\kappa'^2}\alpha_1 + \mathcal{O}(\kappa^2/v_R^2).\\
\end{equation}
After imposing this constraint, $\mu_1^2$ can be as small as \order{\kappa^2}. So just like in the 2HDM case, one of the $\mu^2$ parameters will be much smaller than the other ones. And just like in the 2HDM case, this cancellation does not have an influence on the BG measure, as can be seen in Figure \ref{fig:MuCancellation_LRSM}.

The reason that we do not see any signs of this cancellation is that the relevant terms in the mass matrix all look like $\mathcal{M}^2 \sim \alpha_1 v_R^2 - 2 \mu_1^2$. So when we fill in the formula for $\mu_1^2$ (Eq.\ (\ref{eq:LRminEqn})) the $\alpha_1$ dependence completely drops out of the equation, and the cancellation between $\alpha_1$ and $\alpha_3$ is no longer visible. The same happens when introducing fine-tuning in one of the other $\mu^2$ relations. So fine-tuning in any of the mass parameters of the potential does not show up in the BG measure.

Now we will see what happens when we introduce a cancellation in the seesaw relation. Like argued before, when we solve this relation for one of the $\beta_i$ parameters, we do not get any fine-tuning. However, when there is some numerical coincidence such that the two other $\beta_j$ parameters add up to a very small contribution to $\beta_i$, there might still be fine-tuning present.

When solving the seesaw relation for $\beta_1$ (Eq.\ (\ref{eq:seesaw_beta})), and we have:
\begin{equation}
\beta_3 \approx -\frac{\kappa^2}{\kappa'^2}\beta_2,
\end{equation}
the value of $\beta_1$ will be much smaller than the contributions of $\beta_{2,3}$, and the Dekens measure will give a large value. Once again the BG measure remains small. We obtained a figure very similar to Figure \ref{fig:MuCancellation_LRSM} in this case.

One may wonder whether the large Dekens measure due to large cancellations in the parameters, could be avoided by solving the minimum equations in yet another way. This is indeed possible, but requires selection of some parameters to be unnaturally small. In our numerical analysis we have not allowed unnaturally small parameters by considering the range $[0.1,10]$ as ${\cal O}(1)$. Finding a large Dekens measure may thus indeed indicate large fine-tuning if one requires naturalness and perturbativity of the parameters. The same reasoning holds in our discussion of a large Dekens measure in the 2HDM (section \ref{sec:LargeDekens2HDM}).

\section{Conclusions}
We have investigated two different measures of fine-tuning in two beyond the Standard Model theories. This has shown that these two measures can capture different aspects of fine-tuning. The Dekens measure looks at the minimum of the potential, whereas the BG measure captures features of the second derivative of the potential. Therefore, we argue that both the Dekens measure and the Barbieri-Giudice measure should be evaluated when determining the amount of fine-tuning present in a theory. Of course, we do not exclude that there may be other more effective measures of fine-tuning or there may be other observables besides masses that could be sensitive to fine-tuning.

Furthermore, we conclude that contrary to claims in the literature (e.g.\ \cite{Dekens:2014ina, Deshpande:1990ip, Umezawa:1997yi, Barenboim:1998ib}), it is very well possible to have a large hierarchy of scales in theories like the left-right symmetric models without having fine-tuning, at least at tree-level. This conclusion is reached based on our argument that if it is possible to solve the minimum equations in a way that results in a low amount of fine-tuning, it means that the theory is not fine-tuned in any region of the parameter space, where the parameters are both natural and perturbative.  

Concerning fine-tuning in left-right symmetric models, we conclude that although in a generic LRSM there is no tree-level fine-tuning, the BG measure indicates that imposing the mass of the lightest Higgs boson to be $m_h = 125$ GeV in order to have a phenomenologically viable LRSM may require a moderate amount of fine-tuning.

\textbf{Acknowledgements}
We thank Wouter Dekens, Julian Heeck and Melissa van Beekveld for useful discussions. 
This work has in part been financially supported by the NWO programme ``Higgs as a probe and portal".

\end{document}